# PHYSICAL BASIS OF SUSTAINABLE DEVELOPMENT
# - A FENNTARTHATÓ FEJLŐDÉS FIZIKAI ALAPJAI

László Pál CSERNAI, István PAPP, Susanne Flø SPINNANGR and Yi-Long XIE

**Abstract**
*This work is based on the talk given by Csernai at the Blue Sky International Conference in the Buda Castle on October 30, 2015, Budapest, Hungary. The human development on the Earth is analysed based on basic physical principles and the available resources. The areal and material resources are obviously finite, but the very fundamental energy resources are sufficient for solid and sustainable continuing development. These energy resources can compensate many of the constraints arising from the finite material resources. The development is going in the direction of increasing complexity on the surface of the Earth, due to the increasing green mass and the developing biological and material complex structures. This sustainable development is enabled by the astrophysical conditions and constraints and these conditions provide a good possibility for continuous further development in a sustainable way. This development is characterized by the increasing neg-entropy on the surface of the Earth.*

**Összefoglaló**
*A cikk Csernai László előadásán alapul, amit a Budai Várban a Blue Sky nemzetközi tanácskozáson tartott, 2015. október 30-án. A Földön az emberi fejlődést az alapvető fizikai elvek és a rendelkezésre álló források alapján tanulmányozzuk. A rendelkezésünkre álló terület és az anyagi források nyilvánvalóan végesek, de az alapvető energiaforrások elegendőek egy szilárd és fenntartható folytonos fejlődéshez. Ezek az energiaforrások kiegyenlíteni tudják a véges anyagi forrásokból eredő megszorításokat. A fejlődés az összetettebb rendszerek irányába történik a Föld felszínén, a növekvő zöldtömeg és a fejlődő biológiai és összetett anyagi rendszerek kialakulásával. Ezt a fenntartható fejlődést az asztrofizikai feltételek és korlátok teszik lehetővé, amelyek jó feltételeket biztosítanak a további folytonos fejlődésnek egy fenntartható módon. Ez a fejlődés a Föld felületén a növekvő neg-entrópia kialakulásával jellemezhető.*

**Kulcsszavak**: energia, entrópia, komplexitás, források

I.  INTRODUCTION

The Limits to Growth is a 1972 book, commissioned by the Club of Rome, about the exponential economic and population growth with finite resource supplies (Meadows DH et al. 1972). The book used a computer model to simulate the consequences of interactions between the Earth's resources and human systems. The authors of 'Limits to Growth' published updates in 1992 and 2004.

The original version presented a model based on exponential growth of world population, industrialization. These conditions led to a collapse and reversal of the growth around 2030.
By now we see that these model predictions were unrealistic, and the 2030 as the limit of the growth of human society is unreasonable. See Prof. Norbert Kroó's talk (Kroó, 2015) at this meeting.

The population increase of the Earth is not linear or exponential. Clearly, there are periods with more rapid increase and saturation of the population. These are actually not so much connected the availability of certain resources but the development of technology. The first rapid increase of growth was caused by the development of agriculture a few thousand years ago, then the next is the development of industry that started a couple of hundred years ago. This showed that the availability of energy is just as important for development as the availability of food.

A.  The Availability of Energy

The problem of the Heat Death of the Universe that can arise from the non-decreasing entropy of a closed, near equilibrium system was brought up already by Kelvin in 1852. However, according to our present knowledge, the Universe is not in equilibrium, it is expanding and the expansion changes due to the gravitation.
The heat death idea was also brought up for the Earth, but the Earth is also an open system, it exchanges energy with the surrounding: Enormous radiated energy is received from the Sun, $dQ_{Sun}$, and it radiates in infrared into the universe, $dQ_{Earth}$.

The incoming and outgoing radiations are nearly equal; this is evidenced by the existence of all three phases of the water (steam, water, ice) on the Earth. As we will see later the imbalance is small, it is about $dQ_{Sun} - dQ_{Earth} = +0.6 \text{W/m}^2$, while the maximum of the Solar irradiation exceeds 1000 W/m$^2$. Consequently, we can use the approximation

$$dQ_{Sun} \approx dQ_{Earth} . \qquad (1)$$

At the same time the Sun's radiation and the Earth's radiation is different, these are thermal radiations, but can be characterized by very different temperatures:

$$T_{Sun} \approx 6000\text{K}, \quad T_{Earth} \approx 300\text{K}. \tag{2}$$

We perceive these radiations by their color. Our light-bulbs are usually radiating a color corresponding to 3000 K, and a 4000 K light-bulb is already looking blueish.

Having estimated the heat fluxes and the temperatures of the incoming and outgoing radiation, we can also determine the incoming and outgoing *entropy* currents using the definition:

$$dS = \frac{dQ}{T}. \tag{3}$$

Here we assume locally equilibrated systems, which can be characterized by "intensive" thermodynamic parameters as temperature, $T$, pressure, $p$, etc. In a closed system spontaneous changes must lead to an increase of the entropy, $dS = dQ/T_T > 0$. This thermodynamic entropy is sometimes also called Gibbs entropy. The entropy is an additive ("extensive") quantity, thus we can calculate the change of the Earth's entropy, assuming that both the source, the Sun, and the Earth is in close to local equilibrium, and the energy transfer can be characterized by the temperature of the source:

$$dS_{Earth} = \frac{dQ}{T_{Sun}} - \frac{dQ}{T_{Earth}} < 0. \tag{4}$$

That is the entropy of the Earth is decreasing. What is this entropy and what does it mean that the entropy of the Earth is decreasing. We have seen that the entropy increase was perceived as the Heat Death, so all materials burn and will form structure less dust and smoke. The decreasing entropy (also called as neg-entropy) should be the opposite, but how can we quantify this?

The decrease of entropy may happen if the matter is organized into more complex molecules, living cells, organisms coded with a DNA, and even with the structure of the human brain. These complex living systems are not in equilibrium, these are in change and in development; so their entropy, should not be characterized by a temperature.
Another competing factor is the rather turbulent strong currents in the atmosphere and at the surface of the sea caused by the large temperature differences. The viscosity of air and water damps these currents, while generating entropy increase (Liu et al. 2011).

These considerations indicate that to discuss the development on the Earth should be based on quality and not quantity, and therefore the role of entropy is fundamental in discussing the limits of growth.

## II. GIBBS ENTROPY, SHANNON ENTROPY AND THE ENTROPY OF LIFE

It was observed by Boltzmann that the entropy density of a gas out of equilibrium can be characterized in the space, x, and momentum, p, space, by the entropy density

$$s(x) = -\int d^3 p\, f(x,p)[\ln f(x,p) - 1], \tag{5}$$

where *f(x, p)* is the *(x, p)*-phase space density distribution of the constituent particles of the gas, i.e. the probability that a particle is in a given phase space "volume" element.
(Here we used the convention that the Boltzmann constant is, $k_B = 1$, $c = 1$, and $\hbar c = 1$.) We have to integrate this distribution to all possible phase space volume elements. If we want to receive exactly the same value for the entropy as in thermodynamics, then we have to quantize the volume of the phase space volume elements, based on the uncertainty principle (i.e. that the position and momentum of a particle cannot be determined exactly at the same time). The last term, "-1" is there to secure that the exact low temperature limit of this entropy is the same as in the thermodynamic exact definition. It can be sown that the entropy defined this way returns exactly the same entropy as defined in thermodynamics (Csernai 1994). Boltzmann has shown that this non-equilibrium entropy increases in closed systems until one does not reach the thermal equilibrium, this is described by the "Boltzmann H-theorem". This development actually leads to an increasing 'disorder" in our system.

The information entropy, also called 'Shannon Entropy', was introduced in the mid-1900s [Shannon 1948]. If we assume *n(x)* particles of the same type in a volume element, the entropy expression, eq. (5), takes the form

$$s(x) = -n(x) \sum_i p_i \ln p_i, \tag{6}$$

where $p_i$ is the probability of having a single particle in a given phase-space volume element, *i*, and we have to sum up the contributions of all particles. We can also consider several different objects, with many different states, *i*, for each of them and then the total entropy of this system can be described as

$$S = -\sum_i p_i \ln p_i, \tag{7}$$

where the summation runs over all objects and all of their states.

This way we can for example estimate the human brain's entropy. In ref. (Pénzes et al. 1980), the entropy of humans (as well as some animals) were estimated. The total entropy of the Earth considering all complex systems and life-forms is difficult to calculate precisely, but we can compare the entropy of the different species at their maximum level of complexity, as well as the rate of increase of their entropy during their lifetime. This can be done based on their metabolism and body weight. It is shown that Maximal neg-entropy and the rate of entropy increase provides a good estimate for the life-span of the different species.
The same way one can estimate the increase of neg-entropy of the Earth by considering the increase of population, the increase of the populations of the different species and the increase of the green mass. This is somewhat compensated by the weakly increasing temperature and thus increasing entropy of the atmosphere and the surface layer of the Earth. Nevertheless, the growth of complexity has to dominate the entropy increase from the warming, due to the overall decrease of the entropy of the Earth.

Nevertheless, the decreasing Entropy of the Earth alone is not sufficient to explain the existence and development of life on the Earth.
The neighboring planets have similar entropy imbalance and we still did not see the development of life there. The Earth has a special advantage: the existence of Water in 3 phases, Steam, Water and Ice. This acts as a thermostat, and the latent heat of ice and water vapor establishes a relatively constant temperature environment. This enables the buildup of complex molecules, cells, and life.

In the Sahara, or on the Moon, or Mars, the temperature changes daily by near to 100 °K, so the conditions for a stable gradual and sustainable development are not present.[1]

Let us see how stable is this "Water thermostat" of the Earth.

A.  Ice as the Earth's Thermostat

Nowadays, human beings more and more are occupying the Earth. Due to the hot focus of Global Warming, humans are concerned about the speed of melting the ice. Here we simply estimate the ice melting speed from physical fundamentals, based on some knowledge and observations as follow:

(1) The Earth's ice volume was estimated to be about 29,960,000 km$^3$ [Johnson et al. 2005], i.e.:
$3 \times 10^7$ km$^3$ = $3 \times 10^{16}$ m$^3$ = $3 \times 10^{16} \times 0.9 \times 10^3$ kg = $2.7 \times 10^{19}$ kg.

(2) The Earth surface is: $5.1 \times 10^8$ km$^2$ [Pidwirny 2006], i.e. $5.1 \times 10^{14}$ m$^2$. The area of glaciers on Earth is $1.6 \times 10^7$ km$^2$ [Johnson et al. 2005].

(3) From NASA's observations, there exists on the surface of the Earth a small energy imbalance, $dQ_{Sun} - dQ_{Earth}$, which is measured to be $0.60 \pm 0.17$ W/m$^2$ (Stephens et al. 2012).

---

[1] Still the question arises what happens with the decreasing entropy on other planets where this is not generating complex structures and life forms. On these planets, the lack of water and the thin atmosphere leads to much larger surface temperature differences. This additional structure and the arising strong winds in the thin atmosphere, with turbulent currents lead to significant entropy production balancing the smaller entropy input from the Sun. The radiation outwards is much less uniform compared to the Earth, and a significant part of the Solar irradiation can be directly reflected back. With the rotation of the planet, the absorbed heat still radiated out to the Universe at a much lower temperature. This non-uniform radiation, is proportional to $T^4_{Surface}$, thus it is much higher from the hotter, sunny side than form the dark side of the planet. This way it is acting in the direction of equilibrating the surface temperature, and radiating away the neg-entropy of arising from the surface temperature difference. I.e. the radiation out from such a planet is not thermal, and in eq. (4) incoming and outgoing entropy difference is smaller or negligible because of the reflection and the stronger atmospheric turbulent entropy production.

If the energy imbalance is distributed evenly on the surface of the Earth, then the energy imbalance of the ice surface on the Earth, roughly equals to: $3*10^{20}$ J/year. Then using the latent heat of ice, $3.35 \times 10^5$ J/kg, the melted ice each year is about $3 \times 10^{20}/3.35 \times 10^5$ kg $= 9 \times 10^{14}$ kg, and it will take

$$3 \times 10^5 \text{ years}$$

to **melt all the ice** on the Earth.

Interestingly if we consider all of the ice as fresh water, when the **ice melts in the salty seas,** it will increase the level of the seas due to the density difference between salty and fresh water. Salty water is more dense therefore fresh water cannot displace the amount of salty water equal to its own mass. We can calculate the average rise of sea levels per year. If $9 \times 10^{14}$ kg of ice melts each year, calculating with 1.03 g/ml of salty water density, it will displace roughly $8.7 \times 10^{14}$ kg of seawater. The volume of this water is roughly $8.5 \times 10^2$ km$^3$. By distributing this amount on a surface of $3.6 \times 10^8$ km$^2$ we will obtain a sea level increase of **2.3-2.4 mm per year**.

We also have to take into account that the energy imbalance is heating up the oceans as well. Considering the salinity of a small portion of the ocean, which could absorb the heat, the heat capacity of that portion would be roughly 4100 J kg$^{-1}$ K$^{-1}$ (McDougall et al. 2009) and the energy received on the surface of the ocean ($3.6 \times 10^{14}$ m$^2$) is $6.8 \times 10^{21}$ J/year. With this heat capacity when all the energy is absorbed by that portion, $1.6 \times 10^{18}$ kg ($1.58 \times 10^6$ km$^3$) of sea water would be heated up by 1 K, and if we calculate with a linear thermal expansion coefficient of $2.1 \times 10^{-4}$ K$^{-1}$ the volume of that portion would increase with 330 km$^3$. Distributing this volume on the surface of the sea would approximately result in **0.9 mm increase of sea level per year**.

Adding up the two effects gives nearly exactly the values measured by NASA (Poore et al. 2015).

The estimate of radiated energy imbalance and the observed sea level change are thus consistent. As a matter of fact in the last few thousand years the sea level change was similar, around 1-2 mm per year or 1-2 m per 1000 years. If this sea level change is going to be increased or not is unclear at this moment. IPCC's 2001 projections estimate the range of sea level increase from this value, i.e. **9 cm to 88 cm** for the next 100 years.

In IPCC's worst case scenario this would mean that the ice on the Earth would melt in

$$3 \times 10^4 \text{ years},$$

but by then the Earth's energy supply certainly will not be based on fossil fuels, and until then the ice serves us well for cooling the Earth.

In the industrial age, the amount of energy available for humanity is increasing but again not linearly or exponentially, rather with rapid increases and stagnation periods. These are connected to the takeover of different energy sources, as wood, coal, oil, natural gas, and nuclear energy. All but the last mentioned one are actually converted from the energy of the Sun, with shorter or longer storage or latency periods. In addition, come the more or less direct (renewable) conversions of solar energy via Water, Wind, Photo-Voltaic, energy sources. These forms of energy resources may be limited, particularly the fossil ones, coal, oil and natural gas. These could be available for a limited time of the order of $10^2$ years. The other renewable energy forms have higher economic costs and an intermittent nature, which would require additional expenses as well as additional technological tools for storage.

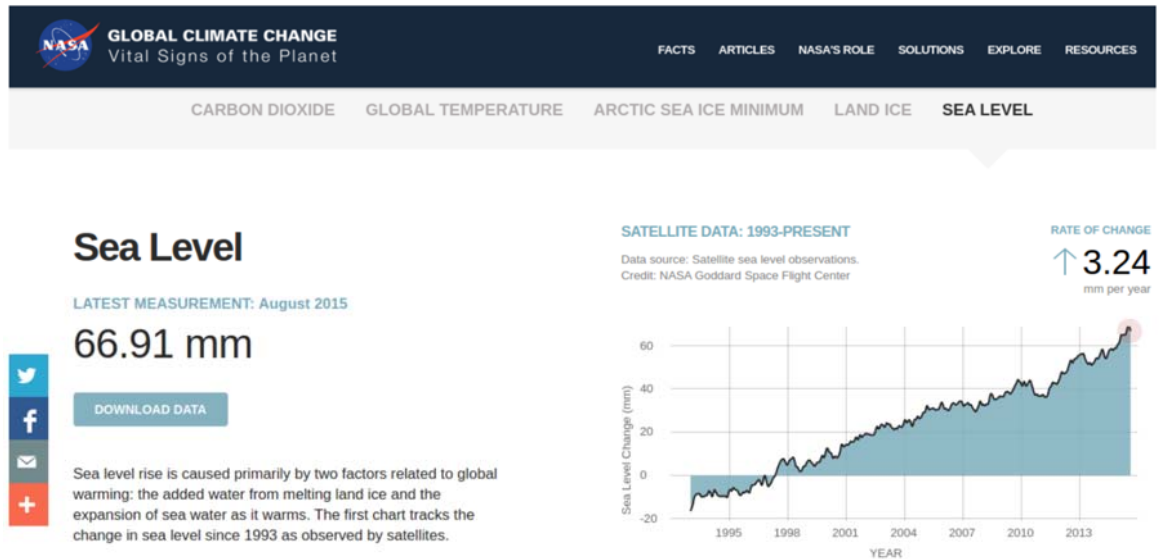

FIG. 1. Measurements of sea level per year according to satellites of NASA. It shows a 3.24 mm increase of sea level per year (Poore et al. 2015).

The Nuclear energy is under development. This is not a conversion of the solar energy for our use, rather the use of those nuclear reactions which occur typically in the stars: fission and fusion. This last form of energy is available at a much larger scale and could last several orders of magnitude longer than the other energy forms dominant up to now. Thus, the energy supply of the Earth is sufficient to Sustainable Development even with still increasing population.

A further hindrance of Growth is the environmental side effects, due to pollution, waste heat and change of the atmosphere, and atmospheric processes due to the emission of climate gases, particles and aerosols. The pollution and waste are actually generating entropy increase, so if our aim is to achieve a sustainable Development measured in quality we have to maximize the development of more complex forms of matter with the least waste production. These side effects are today dominant for the fossil energy production, so these effects will essentially disappear together with the fossil energy.

Still these environmental and atmospheric side effects are at this moment of time important, particularly in rapidly developing part of the World like China.

III.   THE ENERGY MIX OF CHINA AND ITS CONSEQUENCES

China is the most populous country of the World, and the environmental problems related to energy production are there the most severe. Hungary in contrast has relatively modest pollution as nearly half of the electric energy is production is nuclear. About 70 % of the energy production in China is made from coal. This is the most polluting form of energy sources. The question arises what are the present environmental effects of energy conversion in China.

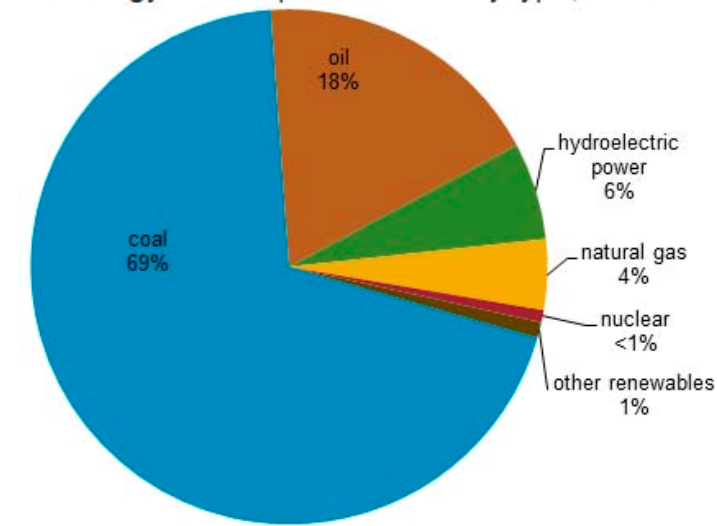

FIG. 2. China's energy production mix in 2011. Source (US EIA 2015).

The question arises does China becomes warmer or colder? Maybe colder due to the smog from energy production?

The present temperature change in China was estimated to be +0.15 °C/ in 10 years (1951-2004) [Ren et al. 2005]. This value is a conservative estimate, i.e. it is a maximum value, and the practical value at different places and times may be lower. The well-known Chinese scientist, Chu Ko-Chen, has studied China's historical climate fluctuations, and his conclusion was that the typical climatic temperature change in China was $0.5 - 1$ °C in every 50-100 years (Chu et al. 1973), but sometimes as much as 0.1 °C per 10 years. The recently measured value of 0.15 °C per 10 years, (Ren et al. 2005) is somewhat greater than 0.1 °C per 10 years, but from Fig. 3 in Ref. (Chu et al. 1973) this changing speed of 0.15 °C/10 years can also be found in China's earlier history.

During the last 30 years, the temperature did have an obvious increase, but it is not unusual, especially considering the urbanization and effect of urban heat islands (Wang et al. 2014).

The temperature during the last 15 years increased in a mild way (Wang et al. 2014, MoEP China 2010), and even in some regions such as in the Beijing-Tianjin-Hebei region, there is a trend of temperature decrease in recent years (MoEP China 2010). This is attributed to the excessive dust and aerosol emission in these regions.

However, according Refs. (Chu et al. 1973, MoEP China 2010), one should also notice that the temperature in China has climbed to the highest level in the last 5000 years. This temperature growth started roughly from 1900.

As a conclusion, China has entered into a hot period since 1950 (Ge et al. 2014), and this increasing of temperature would be maintained for about another 50 years, because the cyclic period of the temperature in China is about 100 years (Ge et al. 2014).

A.     China's Energy resource structure

Thirty years ago, the energy production was almost exclusively based on burning coal and a smaller amount of oil. The energy production increases by more than a factor of 6 in this time and still coal and oil are the dominant sources of energy, but the hydroelectric energy came up to the 3rd place (especially due to constructing the World's largest Hydroelectric power plant of 22.5 GW (equivalent to 22 large Nuclear Power Plants). Just about 25 km downstream on the Yangtze River there is another dam at Yichang City with 2.7 GW power production. Still the total hydroelectric power of China is below 10 %.

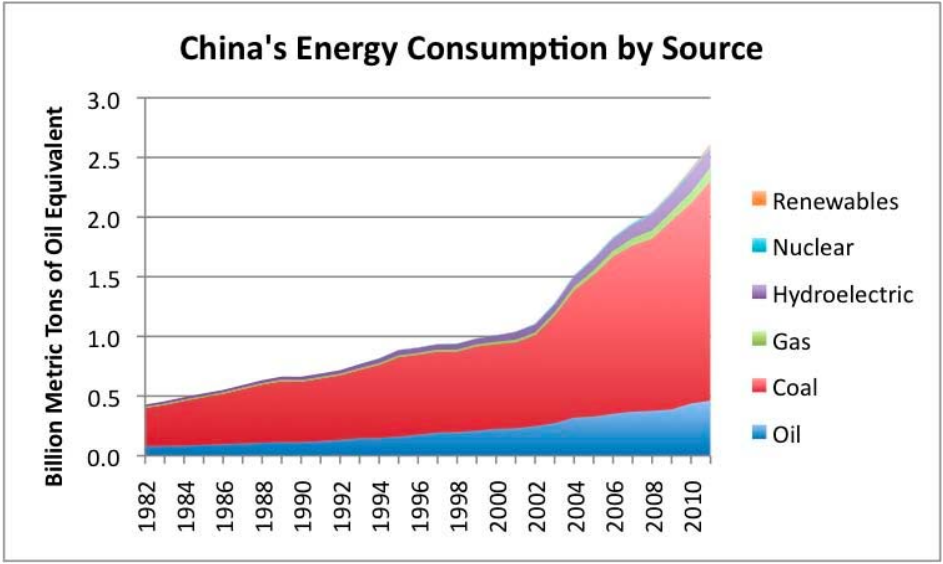

FIG. 3. The development of China's energy production in the last 30 years. Source (Tverberg 2012).

At this time (2014) China has about 20 GWe energy production by Nuclear Power plant and another 26 GW is under construction. By 2020, China plans to reach a fraction of 6 % Nuclear Power in its energy mix.

## IV. LIMITED RESOURCES

Certainly, the surface of the Earth is finite, especially the dry-land, and the population density is increasing. Therefore, the exponential or rapid increase of the population is not sustainable on the long term. Food production is still increasing adequately, but one should consider the land area of one of the most precious resources.

Again, we can take China as example. In the 1960s China had famine, with a population of about 0.7 billion. The one child per family was introduced and strictly enforced. The agriculture is made technologically more intensive with fertilizers, gene modified crops and improved agricultural technologies. Production of fertilizers is highly energy consuming, but this was made possible by increased energy production. In addition, the economic conditions were also modernized; different social productions and different incentives were introduced. This altogether led to a food production, which by 2000 exceeded the demand. This happened without the increase of the agricultural territory but with a significantly higher energy consumption of agriculture. Thus, technological and social advances can relax rigid limits!

The sustainable population growth is also an important question. Developing countries have large reproduction rate and increasing population, while the most developed countries have usually stagnating or slightly decreasing population.

The Chinese one child policy turned out to be unsustainable, as the working Chinese population became too small to cover the needs of the ageing elderly population. This led to changing the one child policy to a two-child policy at the end of 2015. This shows that the issue of sustainability should be regularly reviewed and modified according to the needs. Usually decreasing or lacking resources can be replaced or reproduced, but one should consider the costs and consequences carefully.

This also applies to the questions of energy production, where the proper energy mix, the rate of the change of the mix, and the level of subsidizing or enforcing the change, are of utmost importance. Usually one cannot find or apply a general solution to these problems because these parameters depend on geographical, historical and economic conditions. The balancing the sustainable development among different countries, is a non-trivial questions and may lead to political differences and disagreements (sometimes even wars).

Under these conditions, the sustainable development is a highly complex problem, where natural science and human or social science issues are equally important. Thus, a communication among these different scientific research activities should be much more intensive than earlier.

However, not only the sustainability of the development but the direction of development is changing.

## V. THE DIRECTION OF DEVELOPMENT

As we discussed the Astrophysical conditions of the Earth and the Solar System, secure relatively balanced energy transfer to and from the Earth, stable physical conditions, with physical parameters that enable the sustainable Development of more complex organic molecules, life-forms, and human constructions on all scales. Even developments in the social structures of humanity could be considered as entropy increasing and decreasing changes, although these are difficult to assess quantitatively.

The constrained size of the Earth will not allow unlimited increase of the population, and the amount of material resources will not increase either. These conditions will remain the same for very many years (until population towards other planets will become technically and economically possible and desirable).

The development, already today, goes in the direction of increased complexity of the human life. E.g. cars do not grow but consume less fuel. Even Formula-1 racecars use hybrid technology today. We have much more effective and diverse medicaments, and medical methods. Thus, the exploitation of the increasing neg-entropy is continuing in our present development. Many aspects of these positive developments were mentioned in Norbert Kroo's talk at this conference.

The 70th Anniversary UN General Assembly in September 2015, has uniformly accepted (UN 2015) those goals (27), which serve the peaceful and sustainable development of humanity and the direction of this development.

Regarding the Energy, our Goal (7) is to ensure access to affordable, reliable, sustainable and modern energy for all, peoples and countries. In more detail: (7.1) By 2030, ensure universal access to affordable, reliable and modern energy services; (7.2) By 2030, increase substantially the share of renewable energy in the global energy mix (7.3) By 2030, double the global rate of improvement in energy efficiency.

These goals should be reached by adequate Research, and the solutions should be available and applicable for Developing Countries also.

Let us close these considerations by the thoughts of Nobel Laureate (1978) Alexander R. Todd: The phenomenal rate of change which has characterized our material civilization during this century has been wholly due to the application of scientific discoveries to practical problems - in a word, to science based technology. ... automobiles, television, antibiotics and all the rest - have depended on science. ... Of course, no-one would claim that science has been a wholly unmixed blessing or deny that it has been on occasion misapplied. ... What I wish to argue is that, just as we owe our present civilization and standard of living largely to science, it is only through the further promotion of science and technology that we will find solutions to many seemingly intractable problems ..... (If) we continue to improve our natural knowledge all experience suggests that we will see changes which will radically alter the whole pattern of our lives - or if not of our lives then those of our children and grandchildren; and we shall survive.

ACKNOWLEDGEMENTS

This work is supported by the New Central Europe 2 project. The authors L.P.Cs., I.P. and Y.L.X. thank for the enlightening discussions at the Institute of Social and European Studies in Kőszeg, Hungary.

AUTHORS

**Prof. Dr. László Pál CSERNAI,**
E-mail: laszlo.csernai@uib.no

**M.Sc. István PAPP,**
E-mail: steve.prst@gmail.com

**Ms. Susanne Flø SPINNANGR,**
E-mail: Susanne.Spinnangr@student.uib.no

**M.Sc. Yi-Long XIE,**
E-mail: yi.long.xie.china@gmail.com

Institute of Physics and Technology,
University of Bergen,
Allegaten 55, 5007 Bergen,
Norway